\documentstyle[multicol,aps,epsf]{revtex}
\begin{document}

\sloppy

\title{Phase separation in the two-dimensional electron liquid
in MOSFETs.}

\author{B.Spivak}

\address{Physics Department, University of Washington, Seattle, WA 98195}

\maketitle

\begin{abstract} We show that the existence of an intermediate phase
 between the Fermi 
liquid and the Wigner crystal phases is a generic property of the 
two-dimensional
{\it pure} electron liquid in MOSFET's at zero temperature.
The physical reason for the existence of these phases is a partial 
separation of the uniform phases.
 We discuss properties of these phases and a possible  explanation
of experimental results on transport properties of low density electron
gas in Si MOSFET's. We also argue that in certain range of parameters
 the partial phase separation
corresponds to a supersolid phase discussed in \cite{AndreevLifshitz}.

\end{abstract}


\section{Introduction}

This work is motivated by experiments
\cite{krav,sar,kravHpar,kravHtilt,pudHpar,shah,pudcam,vit,klap1,klap2,okamoto,pudanis,vitkalovParMagn}
 on transport properties of the two dimensional electron system in
high mobility Si-MOSFET's at small electron concentration $n$. These
experiments raised
 doubts about the applicability of the Fermi liquid theory and the
conventional theory of localization \cite{4and,khmelnitskii} to the two
dimensional disordered electron liquid at
low temperatures. 
 The aim of this article 
 is to prove the existence of zero temperature phases of the two
dimensional
 {\it pure} electron liquid in MOSFET's which are intermediate between 
the Fermi liquid and the Wigner crystal.
These phases exist in some interval of concentrations
 $n_{W}<n<n_{L}$. 
The values of the critical concentrations $n_{W}$ and $n_{L}$  are
estimated below.

This phenomenon is due to
a tendency for phase separation which originates from
the existence of a first-order phase transition
between the Fermi liquid and the Wigner crystal phases
as a function of $n$. The difference between the crystal-liquid phase
transition in MOSFET's and the usual
first order phase transitions in neutral systems is the following. In
neutral
 systems with first order 
phase transitions the energy of
the surface between the phases is positive and the minimum of
 the free energy corresponds to 
a minimal surface area and to a global phase
separation.
In charged systems, like electrons on a positive frozen background, 
global phase separation does not occur because of a large Coulomb
 energy associated with a non-uniform distribution of electron density.
The electron liquid in MOSFET's, in a sense, is a system intermediate
between
these 
two limiting cases. Similarly to the neutral systems with 
 first order phase transitions, 
the electron liquid in MOSFET's exhibits phase separation.
On the other hand the surface energy of a minority phase droplet
of a large enough radius 
  turns out to be negative. As a result at
different $n$
 there is a variety of intermediate phases in this system which are
different both from
the Fermi liquid and from the Wigner crystal. 

The electron system with phase separation demonstrates 
a number of unusual features.
If $0<(n-n_{W})\ll (n_L-n_{W})$, the state of the system
 corresponds to a small concentration 
of Fermi liquid droplets embedded into the Wigner
 crystal.
The main difference between such a state and the usual Wigner
crystal
 is that it is not pinned by small disorder and can bypass obstacles,
while the classical crystals at zero temperature are pinned by
an infinitesimally small amount of disorder \cite{larkin}. 
Phenomenologically this state of matter
is similar to the supersolid phase proposed in
 \cite{AndreevLifshitz} for the case of $He^{3}$ 
and $He^{4}$. The difference is that in our case the origin of  
droplets of 
liquid embedded in the crystal is classical electrostatic, whereas, in the
case
\cite{AndreevLifshitz} the existence of vacancies and interstitials in the
ground state of quantum crystals is of quantum origin.

If $0<(n_{L}-n)\ll (n_{L}-n_{W})$, then the state of the
system corresponds
to a small concentration of Wigner
crystal droplets embedded into the
Fermi liquid.
At small concentrations and small temperatures, in principle, these
droplets can be considered as
quasiparticles.

Droplets of a minority phase interact at
large distances via short-range dipole
forces rather than via Coulomb forces. This means that at $T=0$ and at
 small droplet concentration  the system of such
"droplet quasiparticles" should be in
a liquid state similar to $He^{3}$ and $He^{4}$ which are also liquids
 at small densities. Thus we can describe the system by two-fluid
hydrodynamics.
However, the statistics of these quasiparticles
remains unknown.

At zero temperature the one-dimensional boundary between the liquid and
the solid is a quantum object itself. Due to zero-point oscillations of
its position there is a region where the wave function has a form
which is intermediate between the Fermi liquid and the Wigner crystal.  
Since the electron densities of the Wigner crystal and the Fermi
liquid are slightly different the fluctuations of the position of the
boundary is associated with the fact that the number
 of quasiparticles in the Fermi liquid is not conserved.

On the mean-field level this picture of droplet formation in the electron 
liquid in MOSFET's is 
similar to
the partial phase separations which occur in ferromagnetic films
\cite{Doniah}, charged polymers
\cite{polimer,polimer1},
neutron stars \cite{stars},
 doped manganites (see for example \cite{moreo,khomskii}), 
HTC superconductors
 \cite{kiv1,kivelsonemery,kivelsonfradkin,za} and 
 2-dimensional electron systems in the quantum Hall regime 
\cite{shklovsk}.
All these systems demonstrate  a short-ranged tendency for phase separation  
which is thwarted by a long range Coulomb interaction preventing 
global phase separation. 

The paper is organized as follows. In chapter 2 we show that
there is an interval of electron concentrations in which the system is
unstable
with respect to the phase separation. We also estimate  the
size of minority phase droplets   embedded into the majority phase
 and the temperature and  magnetic field dependence of the
 droplet concentration.
In  chapter 3 we discuss transport properties of different nonuniform 
phases associated with the phase separation. In section 4 we compare the
theoretical and experimental results on transport properties of
the low density electron gas in Si MOSFET's.

\section{Phase separation near the point of the Fermi-liquid-Wigner
 crystal phase transition}

In this chapter we show that a partial
 phase separation is a generic 
property of pure 2-D electron liquids in MOSFET's.
Consider a two-dimensional electron liquid of density $n$ in a
 MOSFET separated by a distance $d$ from a metallic gate. Electrons
 interact via Coulomb
interaction while a global electric  neutrality of the system is
enforced by the metallic gate with
a positive charge density $en$. 
The energy density of the system per unit area
$\epsilon(n)=\epsilon^{(C)}+\epsilon^{(el)}$ is a sum of the energy
density of the
capacitor $\epsilon^{(C)}=(en)^{2}/2C$ and the internal energy density of
the
electron
liquid $\epsilon^{(el)}$. In the case of a uniform electron distribution 
 the capacity per unit area is $C=C_{0}=1/d$. 

At high
 electron densities
$na^{2}_{B}\gg 1$
the kinetic energy of electrons is larger than the potential energy and
the
 interaction 
can be taken into account by a perturbation theory. (Here
$a_{B}$ is
the electron Bohr radius.) In this
case the system can be described by Fermi liquid
 theory, the difference between the effective $m^{*}$ and the bare $m$
electron
 masses is small, and $\epsilon^{(el)}=\epsilon^{(el)}_{L}\sim  n^{2}/m$.
On the other hand, in the opposite limit $na^{2}_{B}\ll 1$ (but still $nd^{2}\gg 1$) the
 potential Coulomb energy of electrons is much larger than the kinetic
energy
and the ground state of the system is 
a Wigner crystal with $\epsilon^{(el)}=\epsilon^{(el)}_{W}=-e^{2}n^{3/2}$
(see, for example, \cite{Mott}). Thus at zero temperature there is a critical electron 
concentration $n_{c}$ where the phase transition between the Fermi liquid
and the Wigner crystal phases takes place.  
According to Landau mean field theory this transition is of the first
order (see for example \cite{chaikin}).
The $n$-dependence of the energy densities of the
Fermi liquid $\epsilon_{L}(n)=\epsilon^{C}+\epsilon^{(el)}_{L}$ and the
Wigner
crystal $\epsilon_{W}(n)=\epsilon^{(C)}+\epsilon^{(el)}_{W}$ phases
 near the critical density
$n_{c}$ is shown schematically in Fig.1a.

 In the limit of
 small densities
$nd^{2}\ll 1$, due to the 
existence of the image charges in the gate, the
 interaction between adjacent electrons has a dipole
character.
In this case the ratio between the potential and the kinetic energy decreases as $n$ decreases. Therefore, the small electron $n$ the electron system is a weakly
 interacting Fermi liquid.
Thus we arrive at the conclusion that there exists another critical
 point $n^{(1)}_{c}\sim 1/d^{2}$   which corresponds to a second
 Wigner crystal-Fermi liquid transition. 
The phase diagram of the electron system at $T=0$ is shown in Fig.1b.
If $d< d^{*}\sim 38a_{B}$, than the system is in the liquid state at any value of
$n$. Here the factor $38$ is the result of numerical simulations \cite{cip}.

\subsection{The mean field description of the phase separation.}

In the approximation when $C=C_{0}$ the qualitative picture of the phase
 transition is the same as 
the picture of any first order phase transition in neutral
 systems. In particular, there is an interval
 of electron densities $n_{W}<n<n_{L}$ shown in the Fig.1a where there is
a phase
separation, which means that  there is a spatially
 nonuniform distribution of the Wigner crystal and Fermi liquid phases 
coexisting in equilibrium. In the case of large $d$ one can linearize
$\epsilon^{(el)}_{L,W}(n)$ near the point $n=n_{c}$. As a result, we have 
\begin{equation}
n_{L,W}=n_{c} \pm \frac{(\mu_{W} -\mu_{L})}{2e^{2}d}
\end{equation}
where $\mu_{W,L}=(d \epsilon^{(el)}_{W,L}/d n)|_{n=n_{c}}$.

One can get from Eq.1 an estimate $ n_{c} a_{B}/d $ for the size of
the interval of electron densities 
 where 
the phase separation occurs.
Values of $d/a_{B}$ in various MOSFET's range
from of order one
to 50. 

The relative
 fractions of these phases $x_{W}$ and $x_{L}$ are determined by the
Maxwell rule.
 At $(n_{L}-n)\ll (n_{L}-n_{W})$ the fraction of the area occupied by the
Wigner
crystal $x_{W}\ll 1$
 is small while in the
 case $(n-n_{W})\ll (n_{L}-n_{W})$ the fraction of the area occupied by
the Fermi liquid 
$x_{L}\ll 1$ is small.    
\begin{equation}
x_{W,L}=\pm \frac{n-n_{W,L}}{n_{c}}
\end{equation}

The compressibility of the system $\nu=d^{2} \epsilon /d n^{2} $
should exhibit jumps of order $e^{2}d$ at points $n=n_{L}, n_{W}$.

The crucial difference between first order phase transitions in neutral
systems and in the system of
 electrons in 
MOSFET's arises when one considers shapes of the minority phases.
In the case of neutral systems  the surface energy density $\sigma$
 is positive. Therefore in equilibrium the system
 should have a minimal area  of the surface
separating
the phases, leading to global phase separation. 
On the other hand, in the three dimensional charged systems the global
phase
separation is impossible because of the large Coulomb energy associated 
with the charge separation. It is possible, however, that in this case 
the electron system consists of bubbles and stripes of different
electron density \cite{kiv1,za}, provided the tendency for phase separation
 is strong enough.

The situation in MOSFET's is very different. On one hand, in the
approximation when $C=C_{0}$ global phase separation is possible
 at an arbitrary value of $(\mu_{W}-\mu_{L})$.  On
the other hand, it
turns out that for
 large droplets
of the minority phase the surface energy is {\it negative}. To prove
 this one has to take into account the finite size corrections
to the standard formula for the capacitance \cite{LandauContmed} 
\begin{equation}
C=C_{0}+ \frac{R}{A} \ln \frac{16 \pi R}{ d}
\end{equation}
where $A$ and $d$ are the capacitor area and thickness respectively and
$R=\sqrt{A}$ is the capacitor size.
Consider, for example, the case when $x_{W}\ll 1$.
Then $x_{W}$ can be determined by the Maxwell rule in the approximation when
the second term
 in Eq.3 is neglected and $C=C_{0}$.
Expanding $\epsilon(n)$ with respect to the second term
in Eq.3 and taking
into account 
also the microscopic surface energy we have an expression for the energy
of the 
surface 
\begin{equation}
E_{(surf)}=-\frac{1}{2} N_{W}e^{2}(n_{W}-n_{L})^{2} d^{2}R_{W} \ln \frac{
16 \pi R_{W}}{d} +
N_{W}\sigma
2\pi R_{W}
\end{equation}
 We 
assume
that the Wigner crystal phase embedded into the liquid consists of
droplets of  radius
$R_{W}$ and concentration $N_{W}$ and take into account that inside the
droplet $n\sim n_{W}$.
Thus, at large $R_{W}$ the surface energy Eq.4 turns out to be negative.
We have to find a minimum of Eq.4 at a given total area occupied
by the minority phase, which gives us
the characteristic size of the
droplet 
\begin{equation}
R_{W}\sim \frac{d}{16 \pi} e^\gamma
\end{equation}
with $\gamma=( e^{2}\sigma)/2\pi ({\mu_{W}-\mu_{L}})^{2}$.
A similar expression was obtained in \cite{trug} for a different problem.

The analogous calculation for the case $x_{L}\ll 1$ gives the expression for
the radius of liquid droplets embedded into the crystal which is identical
to Eq.5.

At the point of the transition the values of $\sigma$ and $(\mu_{W}-\mu_{L})^{2}/e^{2}$
are of the
 same order and 
at present nothing is known about the value of the dimensionless
parameter $\gamma$. Even the fact that $\sigma>0$ is not proven.
I would like to also note that in the case of the first order 
phase transitions which are close to the second order one we always have 
$\gamma\ll 1$. 

 In this article we assume that $\gamma\geq 1$.
To illustrate the physical meaning of this inequality we consider
  the case when 2-D electron liquid is compensated by a uniformly charged
positive {\it
frozen} background with a charge density $en$. In this case the
Coulomb energy of a droplet associated with the phase separation is,
roughly, $R/d$ times larger than in the MOSFET's case.
 The most dangerous point with respect to the phase separation
instability 
is $n=n_{c}$ (see Fig.1a). For example,
let us compare the energies of
the
uniform liquid state with $n=n_{c}$
and a nonuniform state which contains two droplets embedded into the
liquid. The first droplet is a liquid with electron concentration
$n_{1}=n_{c}+\delta n$, while the second term is a crystal with
electron concentration $n_{2}=n_{c}-\delta n$. Suppose the droplets have
the same radius $R$.
 Linearizing $\epsilon_{L,W}(n)$ with respect to $\delta n$ we
estimate the energy difference $\delta E$ between these two states as
 \begin{equation}
\delta E\sim (\mu_{L}-\mu_{W})\pi R^{2}\delta n +
\frac{(e\delta n \pi R^{2})^{2}}{R} + 2\pi R \sigma
\end{equation}
The first term in Eq.6 corresponds to a decrease of the energy due to the
 phase separation. The second one corresponds to the positive Coulomb
 energy associated with the nonuniform distribution
of the electron density and the third term is the surface energy.
A minimization of Eq.6 with respect to $\delta n$ gives us  $\delta n\sim
(\mu_{W}-\mu_{L})/R e^{2}$ and
\begin{equation}
\delta E_{min}\sim (2\pi\sigma-\frac{(\mu_{W}-\mu_{L})^{2}}{e^{2}})R
\end{equation}
The assumption $\gamma>1$ means that  $E_{min}$ in
 Eq.7 is positive and that 2-d electron liquid on a
{\it frozen} positive
background does not exhibit a phase separation. (We would like to metion 
that if the value of $|\mu_{W}-\mu_{L}|$ is big enough a microscopic phase
separation in the charged liquid on frozen positive background can take
place even in 3D case. Such situation has been
considered in \cite{kiv1,kivelsonemery,kivelsonfradkin,za,DiCastro})

On the mean field level our problem is similar to \cite{Doniah,polimer}.
Using this analogy we conclude that in the middle of the interval
$(n_{W}, n_{L})$ there is a stripe phase. The phase diagram of the system
is shown schematically in Fig.2. The main difference with
\cite{Doniah,polimer} is the following.
In \cite{Doniah,polimer} all phase transitions between uniform, bubble and
stripe phases are of the 
first order, whereas in our case the transitions between uniform (Fermi liquid
 and Wigner crystal) phases and the bubble phases are continuous. The
 transitions between the bubble phases 
and the stripe phase would be the first order one. However, such a
 transition would have an
interval of concentrations where phase separation would take place.
 In this case the presented above 
arguments could be repeated. Thus we expect more complicated structures than
  bubbles and stripes phase to exist between the bubble and the stripe phases.
 Since the complete solution of this problem remains to be found we
 indicated this  in Fig.2 by shaded lines.  

Let us now estimate dependence
of $x_{W,L}(T, H_{\|})$ on
the temperature $T$  and  the magnetic
 field $H_{\|}$ parallel to the film. It is determined by
the corresponding dependence 
of the free energies for the Fermi liquid and Wigner crystal phases.
At small $T$ and $H_{\|}$ one can neglect the $T$ and $H_{\|}$
 dependences of $\epsilon_{W,L}$ and we have the following expression for the free energy
densities
 of the liquid and the Wigner 
crystal phases 
\begin{equation}
F_{W,L}(H_{\|})=\epsilon_{W,L}-M_{W,L}H_{\|}n-T n S_{W,L}
\end{equation}
where $S_{W}$ and $S_{L}$ are the entropies of the crystal and the liquid
phases respectively,  while $M_{W}$
and  $M_{L}$ are the corresponding spin magnetizations  per
electron.

 As a result, 
one can obtain how $x_{W,L}(T, H_{\|})$, and $n_{cW,L}(T, H_{\|})$ depend
on  $T$ and $H_{\|}$ 
by making the
 following substitution in Eqs.1,2
\begin{equation}
(\mu_{W}-\mu_{L}) \rightarrow
(\mu_{W}-\mu_{L})-(M_{W}-M_{L})H_{\|}-T(S_{W}-S_{L})
\end{equation}
At small $\mu_{B} H_{\|}\ll T\ll E_{F}$ we have $M_{W,L}=\chi_{W,L}
H_{\|}$,
where $\chi_{W}$ and $\chi_{L}$ are linear susceptibilities of the
crystal and the liquid respectively.  (Here $\mu_{B}$ is the Bohr magneton
and
$E_{F}$ is the Fermi energy.)
At low temperature $T\ll E_{F}$ the spin susceptibility of the Wigner
crystal $\chi_{W}\sim \mu_{B}^{2}/T\gg \chi_{L}$ is much
larger than the spin
susceptibility of the Fermi liquid. The entropy  of the crystal
$S_{W}\sim \ln2 \gg S_{L}\sim T/E_{F}$ is mainly due to the spin degrees
of freedom and much larger than the
entropy of the Fermi liquid. 
Thus  $x_{W}$ increases linearly with $T$ and quadratically
 with $H_{\|}$, which means that both
the temperature and the magnetic field parallel to the film 
  drive the
electron 
system toward the crystallization \cite{nosier}.
(We assumed that the temperature is larger than the exchange energy
between spins in the Wigner crystal). 
 These effects are known in the physics of $ He^{3}$ as Pomeranchuk
effects.

In the intermediate interval of magnetic fields $T< \mu_{B} H_{\|} <E_{F}$
spins in the Wigner crystal are completely polarized while the Fermi
liquid is still in the linear regime. In this case $x_{W}$ increases linearly
with $H_{\|}$.

At high magnetic field  $H_{\|}>H^{c}_{\|}\sim E_{F}/\mu_{B}$ both Fermi
liquid and
 Wigner crystal are spin polarized 
and  $x_{W}(T,H_{\|})$ saturates as a function of $H_{\|}$. 
We assume that $\epsilon_{L}(H_{\|}=0)<\epsilon_{L}(H>H^{c}_{\|})$ and,
 therefore
$x_{W}(H_{\|}=0)<x_{W}(H>H^{c}_{\|})$.
On the other hand,
the spin entropy of the Wigner crystal  is frozen in this case. As a
result, at $H_{\|} > H^{c}_{\|}$ the temperature dependence
 of $x_{W,L}(T, H_{\|})$ is suppressed significantly.

\subsection{ Quantum properties of the
droplets of minority phase embedded into the majority one.}

In principle, at small enough concentrations and at small temperatures 
 droplets of the minority phase
embedded into the majority one should behave as quasiparticles. Since the
system is
 translationally invariant,
they should be characterized by momentum (or by quasi-momentum). The
momentum coincides with the flux of mass. Thus
 these quasiparticles
carry  a mass  $M^{*}$, a charge  $eM^{*}/m$ and a spin.
The characteristic temperature of quantum degeneracy is
 $T^{*}\sim N_{W}/ M^{*}$.

The value of $M^{*}$ depends on the mechanism of motion of the droplets,
which in turn depends on whether the surface between the crystal and the
liquid is rough or smooth. 

Consider for example the case of Wigner crystal droplet embedded into
the liquid. In the case of a smooth surface, motion of the
droplet is associated with a redistribution of the liquid
mass on the distance of order $R_{W}$. In this case we can estimate
the effective
mass of the droplet as
\begin{equation}
M^{*}\sim m n_{c}\pi R_{W}^{2}
\end{equation}
In the case of rough surfaces the motion of the droplet is 
associated with melting and crystallization of different parts of it.
Since the $(n_{L}-n_{W})\ll n_{C}$ the liquid mass to be distributed and,
consequently, 
effective mass of the droplet 
\begin{equation}
M^{*}\sim m (n_{L}-n_{W})\pi R^{2}_{W}
\end{equation}
in this case is much smaller than Eq.10.

Droplets of the minority phase
interact at
large distances via short-range dipole
forces rather than via Coulomb forces.
At small enough concentration of the droplets the amplitude of quantum
(or classical) fluctuations of their positions
is larger than the typical distance between them. Thus the liquid droplets
are distributed uniformly over the whole crystal. 
In other words, at $T=0$ the system of such
"droplet quasiparticles" should be in 
a liquid state similar to $He^{3}$ and $He^{4}$ which are also liquids
 at small densities. Thus we can describe the system by two-fluid
hydrodynamics. 
In this case statistics of the "droplet quasiparticles" becomes important.   
 In this respect we would like to mention a difference between
the droplets of the liquid embedded into the crystal and the droplets of 
the Wigner crystal embedded into the liquid.

a. The droplets of the liquid are topological objects which, in 
principle, are not different from vacancies or interstitials in quantum
crystals $He^{3}$ and $He^{4}$.
In order to create such objects in Wigner crystal one has to
add
or to remove from the lattice an integer number of electrons.
Therefore, the liquid droplets have a definite statistics: they are 
either fermions or bosons \cite{AndreevLifshitz}.

 The
main feature of the phase where there are droplets of liquid embedded into
the crystal (supersolid) is it's ability
 to bypass static obstacles. In other words, unlike
conventional crystals
supersolids  are not pinned by disordered potential of small amplitude.
This will manifest itself in the finite conductivity of the system.

From the phenomenological point of view this is very similar to the
scenario
of "supersolid" which has been introduced by
 A.F.Andreev and I.M.Lifshitz \cite{AndreevLifshitz}
for quantum crystals of helium near the quantum melting point.
They assumed that the crystals contain zero point defects (vacancies
 or interstitials) in the ground state and therefore
the number of atoms and number of sites in the crystals are different.
The difference with \cite{AndreevLifshitz} is that the origin
 of the negative surface energy Eq.4
is purely
classical. Conversely, following  \cite{AndreevLifshitz} the existence
of point defects in the ground state could be of quantum origin. Namely,
the
kinetic energy of the
point defects can be larger than the energy required for their creation.
 Thus,
the supersolid phase \cite{AndreevLifshitz} can be considered as a
particular
 case of a more general situation of the phase separation when the radius
of
liquid droplets embedded into the
crystal is of order $n^{-2}$. This would mean that the surface energy is
renormalized to a small
(or negative) value. Indications of the existence of such a phase
have been reported in numerical simulations \cite{pichard}.

b. The case of droplets of the Wigner crystal embedded into the Fermi
liquid
is
different because they are not topological objects. In principle such
droplets can contain an additional charge and spin which can be fractional
or even irrational.
A fundamental problem associated with this fact is that  statistics of
 such quasiparticles is {\it unknown}. 

To illustrate this point we consider a process of tunneling between two
states: a state of uniform Fermi liquid and a state when there is one 
crystalline droplet embedded into the Fermi liquid. These two states
have different total electron charge. Thus the tunneling between
these states is associated with a redistribution of this charge to (and
from) the infinity. It is important that the action $S$ associated with
this process in the pure two dimensional case is finite.  
One can estimate it in a way similar to \cite{lee}.
On distances larger than the droplet size $R_{W}$ one can write the action
in terms of the time-dependent electron density $\tilde{n}({\bf r}, t)$
\begin{equation}
S\sim\int d t d {\bf r} \frac{[e\tilde{n}({\bf r}, t)]^{2}}{C_{0}}\sim 
\int dt \frac{(eM^{*})^{2}d}{m^{2}}\frac{1}{R^{2}(t)} 
\end{equation}
Here $t$ is the imaginary time and $M^{*}$ is given by Eq.11. We 
approximate that
 $\tilde{n}({\bf r}, t)\sim M^{*}/m R^{2}(t)$ at $|{\bf r}|< R(t)$
and $\tilde{n}=0$ at $|{\bf r}|> R(t)$. Eq.12 corresponds to the
potential energy contribution to the action. As
usual, 
the contribution from the kinetic energy is of the same order.
Assuming that $R(t)=v_{F}t$ 
we get an estimate $S\sim (eM^{*})^{2}d/m^{2}R_{W}v_{F}$.
Thus, in principle, the wave function of the object is a coherent
superposition of the wave functions of a uniform Fermi liquid and a
Wigner crystal droplet. In this situation it is quite likely that the
additional charge associated with such an object is not an integer.
This is the reason why the nature of the ground state of the system
remains unknown.

The quantum melting of the phases, which are intermediate between the 
bubble and the stripe phases is even more complicated and we leave this 
question for further investigation.

\section{Transport properties of the electron system with droplets
 of a minority phase embedded into the majority one.}

In this section I will consider cases when quantum statistics of the 
system of droplets of the minority phase is not important. 

The electron-electron scattering conserves the total momentum of the
 electron system
and therefore  does not contribute to the resistance of the system.
To estimate it we have to consider the electron system in the
presence of a random elastically scattering potential.

The electron transport picture in the electron liquid with partial
 phase separation is quite rich. In particular, there is a region of
electron concentrations where the hydrodynamics of the electron liquid is
similar to the hydrodynamics of the liquid crystals
\cite{kivelsonfradkin}.
In this paper we consider only cases where either there are
 crystalline droplets of small concentration embedded
 in the liquid ($x_{L}\ll 1$), or there are liquid droplets with
 $x_{W}\ll 1$ embedded into the crystal. In these situations,
 in principle, there are two types of current carriers in the
 system: electron quasiparticles and 
charged droplets of the minority phase. In this article we will ignore the contribution of the
droplet motion to the charge transport. 
To illustrate the possible $T$ and $H$ dependence of the resistance we
 consider below only several limiting cases leaving a detailed
 analysis for future investigation.  

\subsection{The case when crystal droplets of small concentration are
 embedded in the electron liquid.}

Let us consider the case $x_{W}\ll 1$ when crystalline droplets of
 small concentration 
are embedded into a Fermi liquid. We will assume here  that the Wigner crystal droplets are either pinned by a small scattering potential, 
or have a short mean free path. We also assume that otherwise the impurities do not affect the thermodynamic properties of the system.
The contribution to the resistance of the system from the scattering
 of quasiparticles on droplets has the form
\begin{equation}
\rho=\frac{k_{F}}{e^{2}n l_{(e,W)}}
\end{equation}
where $k_{F}$ is the Fermi momentum of the Fermi liquid, 
$l_{(e,W)}=1/N_{W} R_{W}$ is the quasiparticle mean free path, and
$N_{W}=x_{W}/R_{W}^{2}$ is the concentration of droplets of the Wigner
crystal. 
Thus as followes from Eq.1,2,9,13 at small $T$ the resistance
of the electron system increases linearly in $T$. At small $H_{\|}$ it
increases quadratically in $H_{\|}$, while in the intermediate interval
of 
$H_{\|}$ it increases linearly in $H_{\|}$. 
The saturation of the
magnetoresistance as a function of $H_{\|}$  
takes place at
$H_{\|}>H_{\|}^{c}$ when the electron Fermi liquid gets polarized.

At $H_{\|}> H_{\|}^{c}$ the spin
entropy
 of the Wigner
crystal is frozen. Therefore, as it has been discussed, $x_{W}(T)$ and the
resistance of the system do not have a significant
$T$-dependence.

The $H_{\|}$ dependence of the resistance $\rho(H_{\|})$ of the metallic
phase at small
$T$  is shown schematically
in Fig.3a. The $T$ dependences of  $\rho(T)$ at $H_{\|}=0$ and
$H_{\|}>H_{\|}^{c}$ are shown in Fig.3b.

Eventually at high enough temperatures the crystalline droplets melt.
Since at this point $r_{s}\gg 1$ the melting temperature $T_{m}\ll
\Omega_{p}$
is much smaller than the plasma frequency at the wave vector of order of the inverse inter electron distance. Here $r_{S}$ is the ratio between the potential and the 
kinetic energies of electrons.  
 Let us now discuss the $T$-dependence of $\rho(T)$ in this temperature
interval.
 Though in this
case the
liquid is not degenerate, it is  strongly correlated. Therefore the
electron-electron
scattering in
the liquid is very effective and the local
equilibrium is reached in a short time on a spatial scale of order
$n^{-1/2}$.
 As a result, the flow
of the electron liquid near an impurity can be considered in the framework of
hydrodynamics. In the two-dimensional case the moving electron liquid exerts a
force on an
impurity, which is given by the Stokes
formula $F\sim \eta u/\ln((\eta/nua))$, \cite{LandayLifshitzHyd}. Here $u$,
$\eta$ and $a$ are the liquid
hydrodynamic velocity, viscosity of the electron liquid and the impurity radius
respectively. In a system with a finite concentration of impurities
the logarithmic factor in 
the equation for $F$ should be substituted for $\ln(1/aN_{i}^{1/2})$,
where $N_{i}$ is the concentration of impurities. 
Thus the resistance of the electron system has the form \cite{hruska,spivak1} 
\begin{equation}
\rho(T)\sim \frac{N_{i}\eta(T)}{e^{2}n^{2}}
\ln^{-1}\frac{1}{N_{i}^{1/2}a},
\end{equation}

The viscosity of the strongly correlated liquid in the semi-quantum regime 
 has been considered theoretically in \cite{andreev1} 
for the case of liquid $He^{3}$.
It was conjectured that
\begin{equation}
\eta\sim \frac{1}{T}
\end{equation}
We can apply this result to the case of electron liquid in the semi quantum
regime ($T_{m}\ll T\ll \Omega_{p}$) as well.
Thus we arrive at the conclusion that at high temperatures the
 resistance should decrease
inversely proportional to $T$ and that it should have a maximum at $T\sim T_{m}\sim E_{F}$.
It is interesting to note that, as far as I know, the experimental data
 on the $T$-dependence 
of the viscosity of $He^{3}$ in this relatively high temperature region
are unavailable. However, we can look at data for the viscosity
 of $He^{4}$, which in this temperature interval is supposed to be
similar to 
$He^{3}$ \cite{andreev1}. Though the experimental data for $He^{4}$ are in
a reasonable
agreement with Eq.15, we would like to mention that the viscosity
of $He^{4}$ changes  only by a factor of two in the temperature interval
between $E_{F}$ and the evaporation point.

\subsection{Strongly correlated Fermi liquid in the presence of a 
scattering potential.}

Let us consider the case $n>n_{W}$, but $r_{s}\gg 1$. Then  at $E_{F}\ll
\Omega_{p}$ the electron
 system is a strongly correlated Fermi liquid. The main feature of
 such a liquid is that at small 
distances and at small 
(imaginary) times it behaves like a solid \cite{nosier}.  It has been
suggested
 in \cite{spivak}
that the cross-section of quasiparticle scattering on a short ranged
 impurity  with a radius of order $n^{-2}$, is significantly
 enhanced by the electron-electron interaction.
The nature of the enhancement becomes especially clear if we consider the
interval of
 electron densities close to the critical point $ 0<n-n_{L}\ll n_{L}$ and the case
 when the fluctuations of the external potential
have  a relatively small amplitude.
Then
the system can become split into the regions of a Fermi
liquid and a Wigner
crystal.
I would like to mention that the linear $T$ and $H_{\|}$ dependences 
of the resistance mentioned above are generic for strongly 
correlated electron system and are valid in this case as well.

The
fractions of volume occupied by the Fermi liquid and the Wigner
crystal depend on $n$ and therefore the system should exhibit a 
percolation-type zero-temperature metal-insulator transition as 
 $n$ decreases and the area occupied by the Wigner crystal grows.
There is however a significant difference with respect to the percolation
transition,
which originates from the fact that the position of the Wigner crystal-Fermi liquid 
boundary has quantum fluctuations. It is these fluctuations  
which determine the character of the electron transport near the transition point.
The properties of the current carriers in this region are very different from properties of the Fermi liquid quasiparticles. 

The significance of quantum fluctuations  becomes even more clear
 if we consider quantum
 properties of the surface between the Wigner crystal and the Fermi
liquid. 
At zero temperature the surface is a quantum object itself.
There are at least two scenarios for the state of the surface at $T=0$:
it could be quantum smooth or quantum rough. 
In the first case the excitations of the surface are essentially the Ralaigh
surface waves which conserve the charge inside the droplets. 
In the second case there is a new type of excitations at the surface: 
crystallization waves which do not conserve the total charge inside the droplets \cite{AndreevParshin}. 

The problem of quantum roughening has been
  discussed in the framework of the properties of the boundary
 between solid and liquid $He^{3}$ and $He^{4}$ \cite{AndreevParshin}. 
In the case of 2-d surfaces between a 3-d quantum liquid and crystal
 it has been argued that the surface is always quantum smooth 
\cite{DFisher}. At the moment nothing is known about the state
of  the boundary between 2-dimensional liquid and solid. We would
 like to mention, however, that quantum effects 
(including the quantum roughening)
 are more pronounced in  the case considered above because it is
two-dimensional and  because the jump of the electron
 density in this case is small.    
In any case, due to quantum fluctuations, there is
 a region near the boundary whose properties are intermediate between
 the liquid and the
 solid properties. 

In the conclusion of this section we would like to mention that the
linear in $T$ increase of the resistance at small $T$ is a generic
property of the model.

\section{A comparison between predictions of the theory and
 experimental results in Si MOSFET's.}

\subsection{A review of experimental results on Si-MOSFET's.}

In this subsection we present a short list 
of experimental results on the high-mobility 
two-dimensional electron liquid in Si MOSFET's
\cite{krav,sar,kravHpar,kravHtilt,pudHpar,shah,pudcam,vit,klap1,klap2,okamoto,pudanis,vitkalovParMagn}
which seem to be in contradiction with Fermi liquid theory and with the
conventional
single particle localization theory of disordered
2-dimensional conductors~\cite{4and,khmelnitskii}.

A. The electron system exhibits a "transition" as a 
function of $n$ from a 
metallic phase, where the resistance of the system
saturates at low temperatures, to an insulating phase, where the
resistance increases as $T$ 
decreases. 
The value of the critical concentration $n^{(MI)}_{c}$ of the transition depends on the amount of
disorder in the sample and corresponds to $r_{s}=r^{c}_{s}\gg 1$. Here $r_{s}$ is the ratio between 
the electron potential and kinetic energies.

B. At $T=0$ and for the electron concentration sufficiently close 
to the critical one, increasing the magnetic field $H_{\|}$ parallel to the
film 
drives the system toward the insulating
phase~\cite{kravHpar,pudHpar,pudanis}. 
Thus the critical metal-insulator concentration $n_{c}(H_{\|})$ increases with
$H_{\|}$.

 In the metallic phase ($n>n^{MI}(H_{\|}=0)$) and at small $T$ the system 
exhibits a big
positive magnetoresistance as a function of $H_{\|}$.   
This magnetoresistance
saturates at 
$H_{\|}\ge H_{\|}^{c}(n)$ and $\rho(H_{\|}^{c})/\rho(0)\gg 
1$~\cite{kravHpar,pudHpar}.

C. In the metallic phase at $H_{\|}=0$ and $T<E_{F}$ the  resistance
$\rho(T)$ significantly
increases with increasing temperature.  The characteristic value of
$d \ln\rho/d T>E_{F}^{-1}$ at small $T$ is large 
and depends on the value of $n-n_{c}$.

D. If at $H_{\|}>H_{\|}^{c}$ the system is still in the metallic phase 
($n>n_{c}(H_{\|})$), the $T$-dependence of the resistance is much weaker
than in the $H_{\|}=0$ case ~\cite{shah,vit,klap1}.

E. The value of $H^{c}_{\|}$ decreases significantly as $n$ approaches
$n^{MI}_{c}$. 

\subsection{Qualitative explanation of experimental results.}

In this subsection we present a qualitative explanation of the experimental results
\cite{krav,sar,kravHpar,kravHtilt,pudHpar,shah,pudcam,vit,klap1,klap2,okamoto,pudanis,vitkalovParMagn}

A. {\it The existence of the metal-insulator phase transition.}

 The theoretical picture presented above involves a transition between
 the liquid and the crystal 
as a function of $n$. Therefore it can explain qualitatively the
 existence of the metal-insulator transition observed in the experiments. 
Namely, the fractions of volume occupied by the Fermi liquid and the Wigner
crystal depend on $n$ and therefore the system should exhibit a 
percolation-type zero-temperature metal-insulator transition as 
 $n$ decreases and the area occupied by the Wigner crystal grows.
The transition takes place when the Wigner crystal droplets overlap
and block the electron transport through the Fermi liquid area.

The experimental values of $n_{c}^{(MI)}$ correspond to $r_{s}\sim 10-20$.
 At present it is difficult to say
how close this value is to $n_{L}$, or $n_{c}$ in the pure case. The critical
value  for the transition $r_{s,c}=38$ 
\cite{cip} was obtained by numerical simulations. However it can not be applied to the case of 
electrons in Si MOSFET's because of the existence of two almost degenerate electron valleys.
Another reason for possible inapplicability of the results of  \cite{cip} to Si
MOSFET's is 
that the calculations \cite{cip} were restricted to the case of zero
temperature while the experiments
have been performed at temperatures larger than the spin exchange energy in the Wigner crystal.
Thus the Pomeranchuck effect has not been taken into account in
\cite{cip}. Finally  in 
the critical value of $r_{s}$ can be different in the disordered case. 

B.  {\it The positive magnetoresistance of the metallic phase in
 the magnetic field parallel to the film.}

 The large positive magnetoresistance of the metallic phase in the parallel magnetic field  is connected to the fact
that $\chi_{W}\gg \chi_{L}$ and therefore
 the magnetic field parallel to the film drives the
electron
system toward the crystallization \cite{nosier}.(See Eqs.1,2,9,13). 
The magnetoresistance should saturate when $H_{\|}>H_{\|}^{c}$ and the
electron Fermi liquid is polarized.

C. {\it The temperature dependence of the resistance in the metallic phase.} 

The significant increase of the resistance as a function of temperature
can  be explained naturally as a consequence of the Pomeranchuk effect:
 The spin entropy of the Wigner
crystal is larger than the entropy of the Fermi liquid and, therefore,
 the Wigner crystal regions grow with increasing temperature.

At high temperatures  the droplets of crystal melt. 
It follows from Eqs.13,15 that in this temperature range
 the resistance decreases with 
increasing $T$. It is unclear at present whether the experiments support this picture.

D. {\it The temperature dependence of the resistance in the metallic phase at large $H_{\|}$}.

The Pomeranchuk effect disappears when $H_{\|}>H^{c}_{\|}$ and electron
spins
are fully polarized. In this case entropies of both  the liquid and the solid
are much  smaller than the spin entropy of the crystal at $H_{\|}=0$.
This means that in the leading approximation the areas occupied by the
crystal and the liquid are $T$-independent. 
This explains the fact that in the
metallic
state at $H_{\|}>H_{\|}^{c}$ the $T$-dependence of the resistance
 is much smaller than in the case
$H_{\|}=0$ \cite{vit,klap1}. (The ratio $d \rho/d T (H_{\|}=0)/
d \rho/d T (H_{\|}>H_{\|}^{c})$ can be as big as $10^{2}$).

E. {\it The $n$-dependence of $H^{c}_{\|}$.}

 Perhaps the most direct check of the concept of the Fermi liquid 
which is close to crystallization is the measurement of the $n$-dependence
of the magnetic field
$H_{\|}^{c}(n)$ which polarizes the liquid.
In the case of a
non-interacting Fermi liquid $H_{\|}^{c}=H^{c(0)}_{\|}=E_{F}/\mu_{B}$ is a smooth function of $n$.
The problem of the $n$ dependence of the critical magnetic field $H_{\|}^{c}(n)$ in strongly correlated liquids  near the crystallization point and the origin of the strong
enhancement of the spin susceptibility has been discussed
 in the context of the theory of liquid $He^{3}$ \cite{nosier}.
It has been pointed out that there are two different
scenarios for the origin of the significant (factor 15) enhancement of the spin susceptibility 
of $He^{3}$ near the crystallization point.

a)The system is nearly ferromagnetic which means that it is close
to the Stoner instability.  In this case  the linear spin susceptibility $\chi_{L}$ is
 large, but other coefficients $a_{m}$  in the expansion of the energy 
\begin{equation}
\epsilon_{L}=\chi^{-1}_{L}M^{2}+a_{4}M^{4}+..a_{m}M^{m}
\end{equation}
 with respect to the spin magnetization $M$ are not small. Here $m$ is an
even integer.  
In this case $H^{c}_{\|}\sim H_{\|}^{c(0)}$, which is relatively large.

b) The system is nearly solid. In this case both $\chi^{-1}_{L}$ and
 other coefficients $a_{m}$ in 
 the Eq.15 decrease significantly as $n$ approaches
 the crystallization point $n_{c}$. In this case
 $H_{\|}^{c}(n)\ll H_{\|}^{c(0)}$ is small.

In the case of $He^{3}$ the value of $H_{\|}^{c}(n)$ has never been measured.
In the case of electrons in Si-MOSFET's it has been measured 
in \cite{vitkalovParMagn}. A dramatical decrease of $H_{\|}^{c}(n)$
compared to
 $H_{\|}^{c(0)}$ has been
observed as $n$ approaches $n_{c}$. In our opinion these experimental results
support the model of a nearly solid Fermi liquid which is at $r_{s}\gg 1$.
 Conversely  it is unlikely that the system is close to
the Stoner instability.

\subsection{A comparison with alternative explanations
 of experiments on transport properties of the metallic
 phase of the electron system in Si MOSFET's.}

In this section we compare the explanation presented above with
 another explanation 
given in ~\cite{sarma1,dolgopolov,dolgopolov1,sarma,dolgopolov2,Aleiner}.
 It is based on the fact that a single short range impurity 
in a metal creates  Friedel oscillations of the electron density.
Due to the electron-electron interaction the quasiparticles in the
 metal are scattered
not only from the impurity but also from the modulations of the
 electron density.
 At finite temperature 
the Friedel oscillations decay exponentially at distances larger
 than the coherence length of the normal metal $v_{F}/{T}$. 
As a result, at low
temperatures $(\rho(T)-\rho(0))\sim C T$ with $C>0$ 
~\cite{sarma1,dolgopolov,dolgopolov1,sarma}.
The exchange contribution to the resistance has not been taken into
account in \cite{sarma1,dolgopolov,dolgopolov1,sarma}. It has been shown
\cite{Aleiner} that in the presence
of the exchange interaction at  $\tau^{-1}\ll T\ll E_{F}$ the quantity
$(\rho(T)-\rho(0))$ remains linear in $T$.  However, at $r_{s}\ll 1$ the
coefficient $C<0$
 has negative sign, which is different from \cite
{sarma1,dolgopolov,dolgopolov1,sarma}. On the other hand, 
the expriments were performed in the regime $r_{s}>1$. They yeald a
positive value of the coefficient $C>0$. At finite value of
$r_{s}\sim 1$ the theory  \cite{Aleiner} predicts, that the coefficient
 $C$ chenges it's sign agein and become positive.

At relatively high temperatures $T\sim  E_{F}$ one can neglect the
interference corrections and the temperature dependence $\rho(T)$ is
determined by the corresponding dependence of the thermal velocity and
the electron scattering cross-section in a nondegenerate gas. In this
regime $\rho(T)$
decreases with increasing $T$ \cite{sarma1,sarma}. 
At this
point we would like to mention
that at low $T$ Eqs.1,2,9,13 also predict the increase of the
resistance linear in $T$ as well.
Eqs.14,15 also predict the existence of the maximum of $\rho(T)$ at
$T\sim E_{F}$ and decrease of the resistance at $T>E_{F}$.
Thus, both the theory presented above and
~\cite{sarma1,dolgopolov,dolgopolov1,sarma,dolgopolov2,Aleiner}, in
principle,
could explain qualitatively the $T$-dependence of the resistance of the
metallic state.

The situation with the magnetoresistance in the parallel magnetic field is
more delicate.
Strictly
speaking, the
interference corrections to the Drude conductivity calculated in 
~\cite{sarma1,dolgopolov,dolgopolov1,sarma,dolgopolov2,Aleiner}
are relevant only at $r_{s}\ll 1$ and at small $T$ and $H_{\|}$, when the
effects
are small.
On the other hand, at $H_{\|}>H_{\|}^{c}$, when the effects are
large, the interference
corrections  are irrelevant, and the value of the
magnetoresistance $(\rho(H_{\|})-\rho(0))$ is determined by the $H_{\|}$
dependence of the Drude part of the resistance, which is due to the
$H_{\|}$
dependences of the Fermi momentum and the scattering cross-section
of quasiparticles. (This part of the magneto resistance has not been taken
into account in ~\cite{Aleiner}).
In this case a single electron theory  yields a big and negative
 Drude magnetoresistance  in
contradiction with the experimental fact that it is big and positive.

 In connection
 with this I
would like to make several points.

a. At $r_{s}>1$, 
the diagrammatic calculations are not under control.
In the framework of the conventional diagram technique
it is difficult to account for  all effects associated with the
 strong correlations at $r_{s}\gg 1$, including the giant
 renormalization of the electron
scattering
cross-section on impurities, the effects of phase separation,
existence of the crystallization waves at the boundary between the two
phases, and, finally, the Wigner crystallization itself.

b. The mechanism considered in
~\cite{sarma1,dolgopolov,dolgopolov1,sarma,dolgopolov2,Aleiner}
can not explain the increase of the
resistance as a function of $H_{\|}$ and $T$ which
 is significantly larger than unity.
This is because the amplitude of the potential created by the Friedel 
oscillations of the density created by an impurity potential is smaller
 than the impurity potential itself. This theory also can not explain
why the temperature dependence of the resistance $\rho(T)$ is suppressed so dramatically
by the magnetic field parallel to the film.
  On the other hand, the theory presented in this article can explain these facts.  

c. These mechanisms of $T$ and
 $H_{\|}$ dependence of the resistance  are based on very different
physics.
This can be seen, for example, from the fact that all single electron
 interference phenomena including 
the Friedel oscillations are smeared by finite temperature.
 Conversely, the fraction of the Wigner crystal
increases with temperature.  

d. The amplitude of the Friedel
oscillations is suppressed significantly in the case when the scattering
 potential is a smooth function of coordinates on the scale of the
electron
 wavelength. This is exactly what happens when 
the scattering cross-section is significantly renormalized by the
 fact that near a short range impurity there are crystalline
 droplets and the position of the crystalline surface 
exhibits quantum fluctuations. Thus, in a sense, the mechanisms based on
 single electron interference and the mechanism based on quantum
 fluctuations of the solid-liquid boundary compete with each other.  

In order to distinguish between these two mechanisms one needs to perform
 experiments on samples with higher mobility, where effects considered in this
 article will be much larger than unity.

Finally we would like to mention that the theory 
\cite{sarma1,dolgopolov,dolgopolov1,sarma,dolgopolov2,Aleiner} may be
 relevant to experiments on the two dimensional electron system
 in GaAS samples \cite{galium,galium1}.
 
 \section{conclusion}

We have shown that due to the existence of  metallic gates in MOSFET's
 the phase separation is a generic property of pure
 electron liquids. The proof is based only on the assumption about
 the existence
of the first 
order phase transition between the uniform Fermi liquid and the Wigner
crystal phases
 and on electrostatic properties of two dimensional electron system. 
This distinguishes the theory presented above 
 from the theories \cite{KivelsonChakrovarty,spivak1}, which attempted to
 explain the experiments using
 the fact that in the 2-d electron
liquid there is 
 a first order phase transition between the Fermi-liquid and the Wigner
crystal which is destroyed by small disorder. 
This difference, however, manifests itself only at relatively small
 values of $d$ and 
at relatively small amplitude of the disorder.
Qualitative pictures of the $T$ and $H$-dependences of the resistance of
the "metallic" phase are, roughly speaking, the same for the model presented
above and for those considered in \cite{spivak1}.

It is an open question how a disorder of finite amplitude affects the
results presented above. In some regimes the system
 can demonstrate a glassy behavior characteristic for crystals in
 the presence of disorder. Experimental indications of glassy
 behavior of the electronic system in Si MOSFET's have been
 reported in \cite{Gershenson,Popovic}.

In this paper we considered only bubble phases which exist near the
 critical concentrations $n_{L}$ and $n_{W}$.
In the interval $n_{L}<n<n_{W}$ the system, will probably exhibit
 a sequence of quantum phase transitions. In particular, it is likely that
  at electron densities close to $n_{c}$ there is a stripe phase, which
is similar to \cite{kiv1,za,shklovsk}.

In conclusion we would like to mention that the picture presented above 
is
 in many respects similar 
to the quantum critical point of strongly correlated electron systems
considered in  \cite{laughlin,sachdev,millis}. In particular, the
Fermi
liquid state with densities close to $n_{W}$ will demonstrate
 very large sensitivity to
imperfections, which is characteristic for the "almost
critical" quantum state
\cite{laughlin}.

This work was supported by Division of Material Sciences, 
U.S.National Science Foundation under Contract No. DMR-9970999.
We would like to thank A.F.~Andreev, E.~Abrahams,   
S.Chakrovarty, A.Efros, M.Gershenson, S.~Kivelson, S.~Kravchenko,
 L.~Levitov, D.~Maslov, V.~Pudalov,  B.~Shklovskii, M.~Sarachik,
and S.~Vitkalov for useful discussions.

\newpage

\begin{figure}
  \centerline{\epsfxsize=10cm \epsfbox{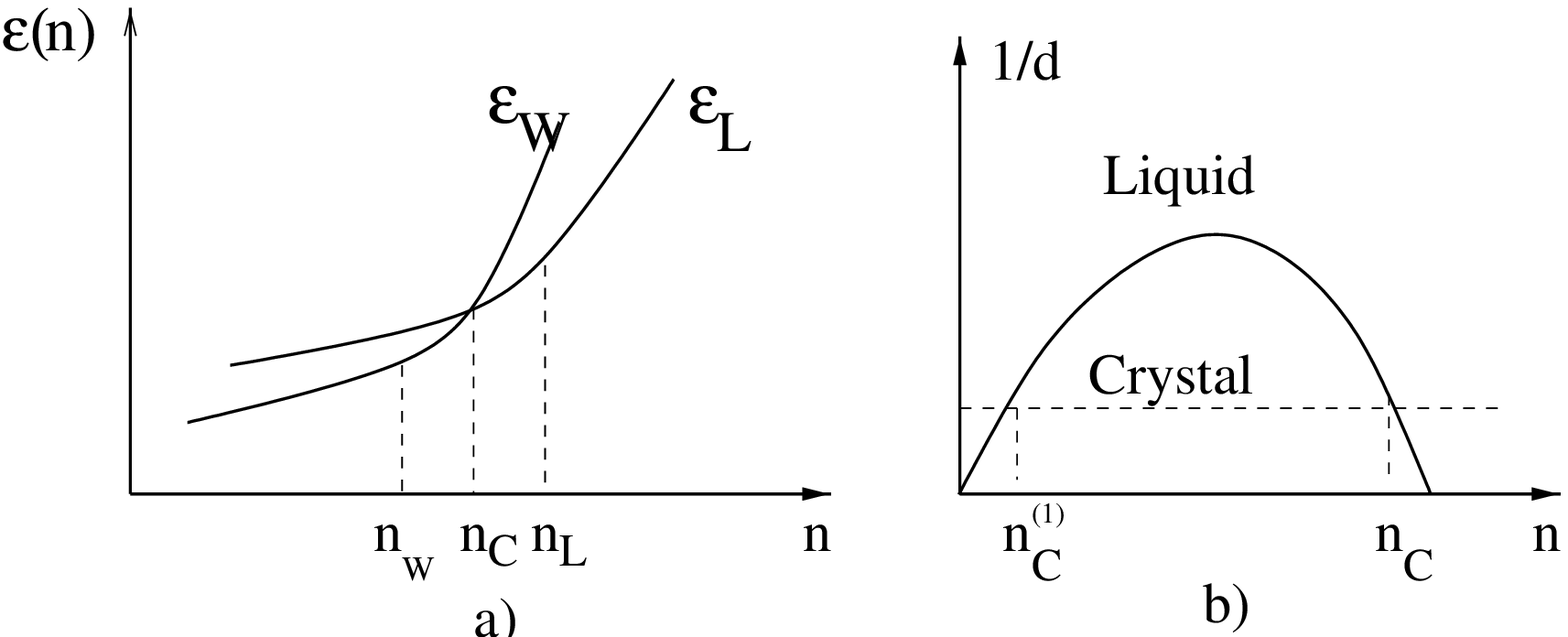}}
  \caption{a. The dependence of the energy densities of 
the Wigner crystal and the Fermi liquid phases $\epsilon_{W,L}(n)$ on the 
electron density $n$. Symbols $W$ and $L$
correspond to the Wigner crystal and the Fermi liquid phases 
respectively. b. The effective phase diagram of the 2D electron system at
zero
temperature.} \
  \label{fig:fig1}
\end{figure}

\begin{figure}
  \centerline{\epsfxsize=10cm \epsfbox{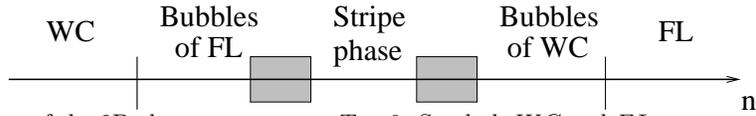}}
  \caption{The phase diagram of the 2D electron system at $T=0$. Symbols
$WC$ and $FL$
correspond to the Wigner crystal and the Fermi liquid phases
respectively. The shaded regions correspond to phases which are more
complicated than the bubble and the stripe phases.} \
  \label{fig:fig1}
\end{figure}

\begin{figure}
  \centerline{\epsfxsize=10cm \epsfbox{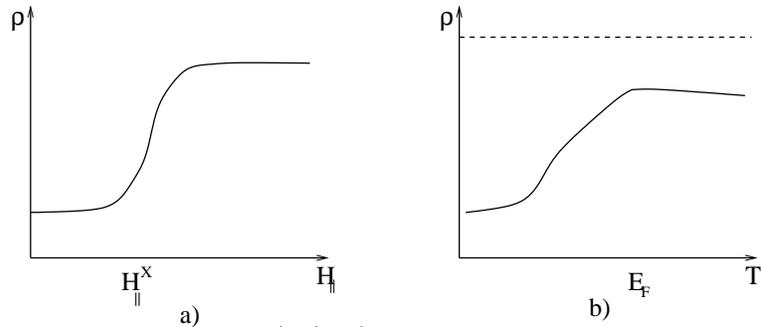}}
  \caption{a)The $H_{\|}$ dependence of the resistance $\rho(H_{\|})$. b)
The temperature
dependence of the resistance $\rho(T)$. The solid line corresponds to the
case
$H_{\|}=0$, while the dashed line corresponds to the case
$H_{\|}>H^{c}_{\|}$ } \
  \label{fig:fig1}
\end{figure}

\end{document}